\documentclass{article}
\usepackage{ltwol2e}
\def \ts{\thinspace}
\def \nts{\negthinspace}
\def \ns{\enspace}
\def \Wg{W\negthinspace g}

\def \GeV{{\rm \enspace GeV}}
\def \TeV{{\rm \enspace TeV}}
\def \beq{\begin{equation}}
\def \eeq{\end{equation}}
\def \beqa{\begin{eqnarray}}
\def \eeqa{\end{eqnarray}}
\def \Aud{{\cal A_{\uparrow\downarrow}}}
\def \half{\hbox{$1\over2$}}
\def \W{{\scriptscriptstyle{W}}}

\def\simge{
    \mathrel{\rlap{\raise 0.511ex 
        \hbox{$>$}}{\lower 0.511ex \hbox{$\sim$}}}}

\def\simle{
    \mathrel{\rlap{\raise 0.511ex 
        \hbox{$<$}}{\lower 0.511ex \hbox{$\sim$}}}}

\def\agt{\simge}                              

\begin{document}


\title{Observing Spin Correlations in Single Top Production
and Decay}

\author{Gregory Mahlon$^{*}$}

\address{Department of Physics, McGill University, 
3600 University St., Montr\'eal, QC  H3A 2T8, Canada\\
Electronic address:  mahlon@physics.mcgill.ca}

\date{\today}

\twocolumn[\maketitle\abstracts{
The weak decay of a polarized top quark has a rich structure
of angular correlations among its decay products.  We describe
some of the noteworthy features of such decays.  
We then examine the three modes of single top production, with
an emphasis on
the status of calculations of the degree and direction
of polarization at Run II of the Tevatron.  
The associated production of a top quark with
a $W$ boson will be very difficult to observe at the Tevatron.
Prospects for observing
the other two production mechanisms are much better.
The spin state of the produced top quark is understood
at leading order in the $W^{*}$ production channel.
For the $\Wg$-fusion process, we present new results including
the resummed
logarithmic corrections of the form $\ln(m_t^2/m_b^2)$.
The spin-dependence of the complete order $\alpha_s$ corrections 
to either of these processes has not yet been determined.
The fraction of spin
up top quarks is large ($\agt 95\%$) for both modes
provided that the appropriate spin quantization axis is chosen.
\hfill McGill/00-33
}]


\section{Introduction}

One of the major physics goals of Run II at the Fermilab
Tevatron will be to study the top quark in as much detail as
possible.  
The anticipated large data sets 
(2 fb$^{-1}$ in Run IIa and 
as much as 15 fb$^{-1}$ in Run IIb) will allow 
for refined measurements of the top quark mass and production
cross sections.  
In addition to $t\bar{t}$
pairs, single $t$ quarks are expected to be produced in
great enough numbers to allow for their observation
and a direct measurement of $\vert V_{tb}\vert$.
Furthermore, it will be possible to study
the kinematic and angular distributions associated with 
various top quark production and decay channels.

A unique feature of single top production is the large net
polarization ($\agt 95\%$) for the appropriate choice
of spin quantization axis.\cite{OptimalBasis} 
The top quark provides the only
laboratory in which we can study the properties of
an isolated quark:
the large value of $m_t$
implies that its weak decay
will take place before the strong interaction
can affect its spin.\cite{Bigi}\ns\ts
Studies of the spin-induced angular
\begin{figure}[h]
\vskip4.5cm
\includegraphics{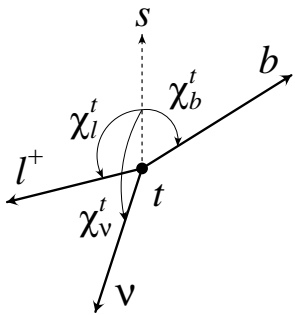}
\caption[]{Definition of the top quark decay angles in the
top quark rest frame.  The direction of the top quark spin
is indicated by the vector $s$.  For simplicity, we have
omitted the $W$ boson and its decay angle 
$\chi_\W^t \equiv \pi-\chi_b^t$.
Although we have drawn this
figure assuming a leptonic $W$ decay, the same correlations
hold in a hadronic decay if we replace the charged lepton by
the $d$-type quark and the neutrino by the $u$-type
quark.}
\label{DecayAngles}
\end{figure}

\noindent
correlations among
the top quark decay products are possible in both the
$t\bar{t}$ and single $t$ production modes.  Although
$t\bar{t}$ pairs will be produced in greater numbers than
single tops, the relative simplicity of the final state
in the single top case  may compensate for the smaller
statistics.  In particular, with only one top quark to reconstruct,
the combinatoric background 
for single $t$ should be smaller 
than for $t\bar{t}$.  
Furthermore, as we shall see below, events
where the $W$ boson decays leptonically have the most distinctive
correlations.   Thus, the largest correlations in $t\bar{t}$
events occur when the kinematics are the most difficult
to constrain because of a pair of invisible neutrinos.
This reconstruction ambiguity is less severe in single
top events, where there is just one neutrino.


\section{Polarized Top Decay}

Within the Standard Model, 
the dominant decay chain of the top quark is
\beq
\includegraphics{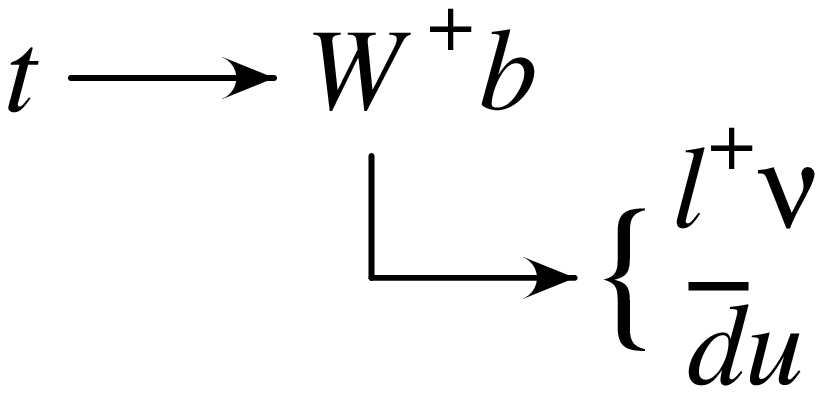}
\phantom{\Bigl[\over\Bigl[}
\eeq
Because of the parity-violating $V-A$ structure of the 
couplings to the $W$ boson,
the decay products of a polarized top quark possess a rich
structure of angular correlations.
For concreteness, we will describe the leptonic $W$ decay.
However, everything which we say about the charged lepton
applies equally to the $d$-type quark when the $W$ decays
hadronically.

The simplest set of correlations are most-easily understood
in the top quark rest frame, where we define 
$\chi^t_i$ to be the angle between the $i$th decay product
and the top quark spin quantization axis (see Fig.~\ref{DecayAngles}).
The decay angular distributions are simply linear
in the cosine of these decay angles:\ts\cite{alphas}
\beq
{1\over\Gamma_T}\thinspace
{ {d\Gamma}\over{d(\cos\chi^t_i)} }
=
{1\over2}
\Bigl( 1+\alpha_i\cos\chi^t_i \Bigr).
\label{dGamma}
\eeq

\begin{figure}[t]
\vskip5.5cm
\includegraphics{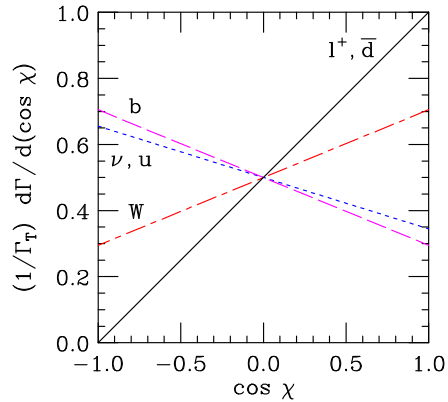}
\caption[]{Angular correlations in the decay of a spin up top
quark.  The lines labeled $\ell^{+}$, $\bar{d}$, $b$, $\nu$,
$u$, and $W$ describe the angle between the spin axis and the particle
in the rest frame of the top quark.}
\label{dGammaPlot}
\end{figure}

\noindent
Eq.~(\ref{dGamma}) applies to spin up $t$ quarks as well as to
spin down $\bar{t}$ quarks.
The analyzing power $\alpha_i$ encodes the degree
to which each decay product is correlated
with the spin of the parent top quark.
Table~\ref{alphaTable} lists the values of the $\alpha_i$'s
for each of the top quark decay products.  The corresponding
angular distributions are plotted in Fig.~\ref{dGammaPlot}.

\begin{table}[t]
\centering%
\caption{Correlation coefficients $\alpha_i$ for both
semileptonic and hadronic top quark decays.\protect\cite{alphas,fns}
The first
two entries are a function of $m_t^2/m_\W^2$, and have
been evaluated using the PDG2000 average values\ts\protect\cite{PDG}
$m_t = 174.3 \pm 5.1 \GeV$ and $m_\W = 80.419 \pm 0.056 \GeV$.
\lower2pt\hbox{\protect\phantom{j}}
\label{alphaTable}}
\begin{tabular}{lcclr}
\multispan5\hrulefill \\[0.05cm]
&Decay Product &\quad& \qquad\quad$\alpha_i$  \\[0.2cm]
\multispan5\hrulefill \\[0.1cm]
\ns\ns\ns\ns\ts
& $W$  &\quad&  $\protect\phantom{-}0.403 \pm 0.025$   
& \ns\ns\ns\ns\ts \\
& $b$  &\quad&  $-0.403 \pm 0.025$  & \\
& $\nu_{\ell}, u,$ or $c$
     &\quad& $-0.324 \pm 0.040$ &\\
& $\bar{\ell}, \bar{d},$ or $\bar{s}$
     &\quad& $\protect\phantom{-}1.000$ (exactly) & \\[0.1cm]
\multispan5\hrulefill 
\end{tabular}
\end{table}

According to Table~\ref{alphaTable}, the charged lepton
is maximally correlated with the
$t$ spin direction, $\alpha_\ell = 1$, independent of the
$t$ and $W$ masses.
Amusingly, the charged lepton possesses a stronger
correlation than its parent, the $W$ boson.
The resolution to this minor mystery lies in the significant
interference between the  polarization
states of the intermediate $W$ boson.\cite{mumu99} 
It is well-known that the $W$ bosons emitted from decaying
top quarks have a specific 
mixture of helicities\ts\footnote{It turns out that the
helicity basis is the only spin basis in which just two of the
three spin states contribute.  Hence, it is the most natural
basis for discussing the $W$ spin in top quark decays.}
in the Standard Model:  the right-handed helicity state
is absent, while the left-handed and longitudinal states
are present in the ratio $2m_\W^2 : m_t^2$ 
(see Eq.~(\ref{Wdist}) below).  
In Fig.~\ref{WleftWlong} we have replotted
the distribution of the charged lepton in the top quark rest frame,
along with the results that would be obtained either by including only
left-handed or longitudinal $W$'s.
By comparing these distributions we see that there is complete
destructive interference between the two $W$ polarization states
when the charged lepton is emitted antiparallel to the $t$
spin.  In the forward direction, however, there is a large degree
of constructive interference.  Thus, not all of the 
information about the $t$ spin possessed by the $W$ bosons is
reflected by their distribution in $\cos\chi_\W^t$:  
additional information
is contained in the $W_{\rm left}$-$W_{\rm long}$ interference 
terms, and that information is imparted to the charged lepton,
endowing it with its maximal analyzing power.
For a spin down top quark, all of the correlations are reversed,
as is the sign of the interference term.   In fact, in the sum
over the two top quark spins, the interference term exactly cancels
point-by-point in phase space.
Thus, for an unpolarized
sample of top quarks the interference term plays no role in the
charged lepton distribution.  

%
\begin{figure}[t]
\vskip5.2cm
\includegraphics{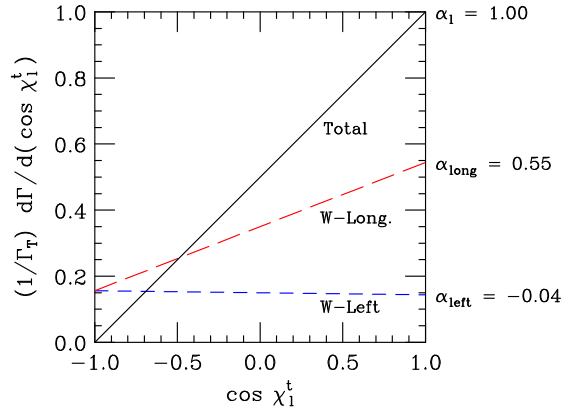}
\caption[]{Angular distribution of the charged lepton in the
top quark rest frame, assuming that only left-handed or longitudinal
$W$ bosons appear in the intermediate state.  The relative areas
under these two lines reflects the ratio of left-handed to longitudinal
$W$ bosons in $t$ decay.
The solid line is the quantum mechanical sum of the two contributions,
and exhibits constructive (total destructive) interference when the 
lepton is emitted parallel (antiparallel) to the $t$ 
spin.\protect\cite{mumu99}
}
\label{WleftWlong}
\end{figure}
%
\begin{figure}[t]
\vskip4.6cm
\includegraphics{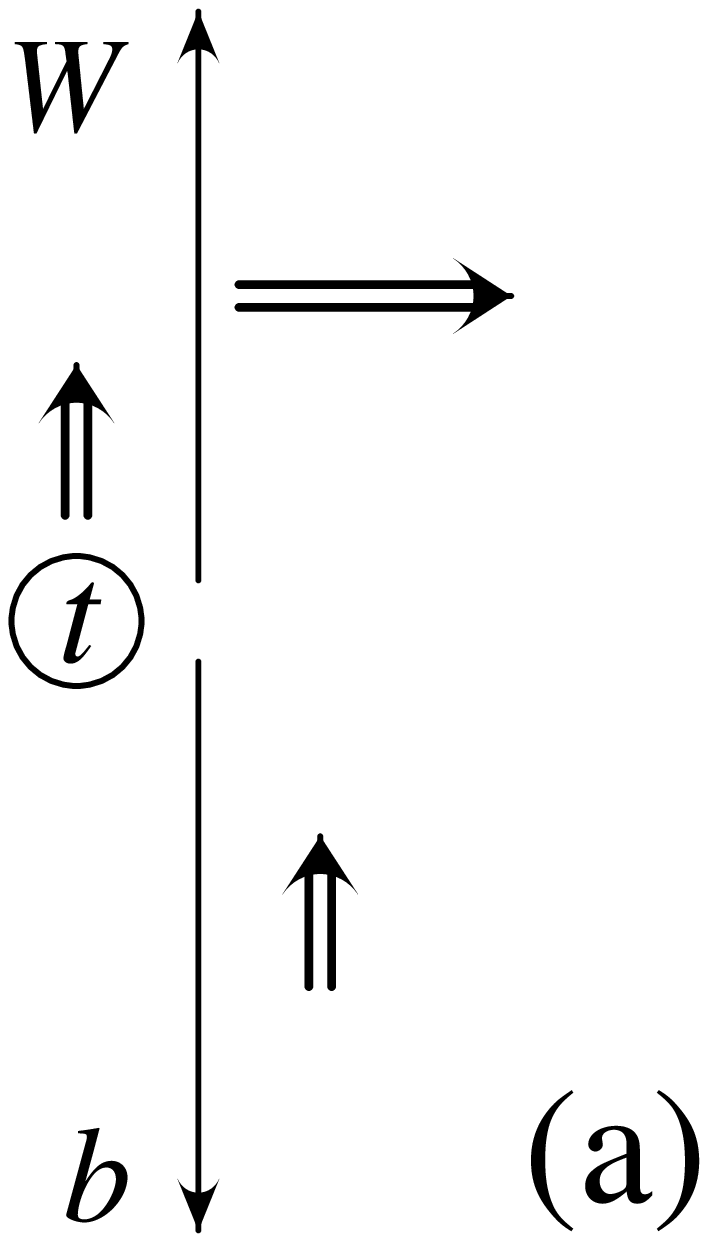}
\includegraphics{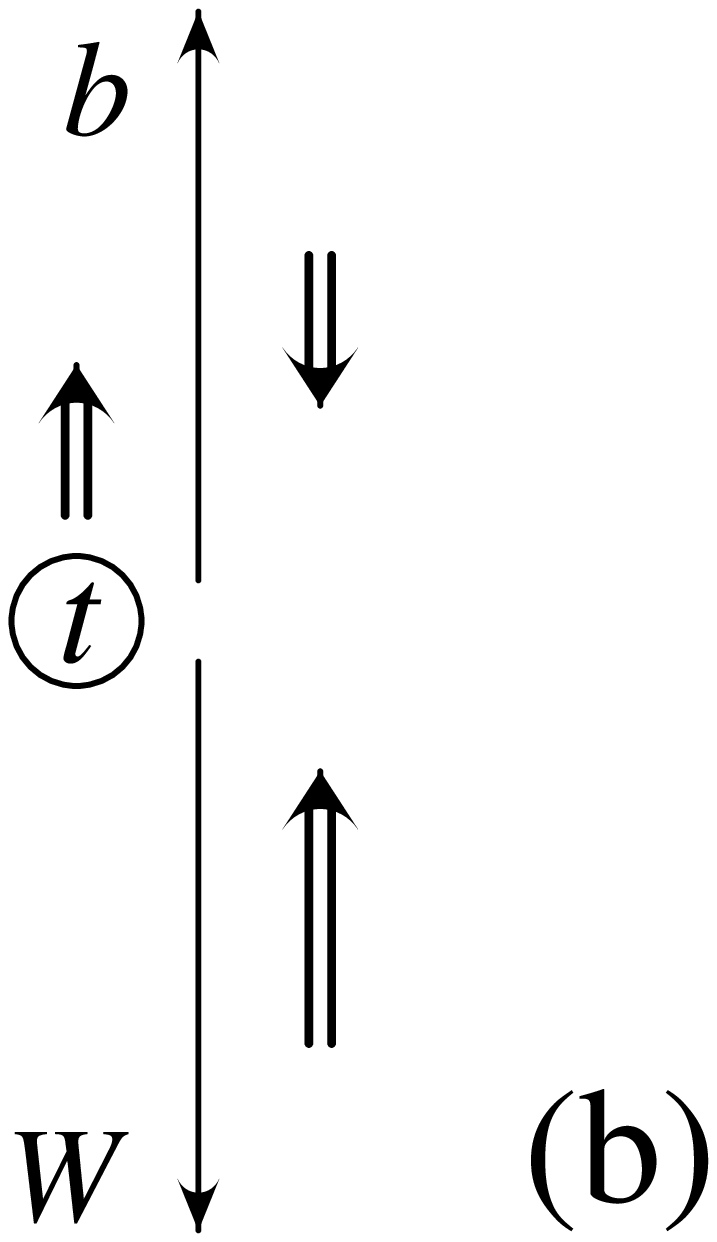}
\caption[]{Angular momentum conservation in the decay of
a polarized top quark.  Since the $b$ quark is effectively
massless and couples to a $W$, it is produced
with left-handed helicity.  (a)  $W$ bosons emitted parallel to the
top quark spin are longitudinal.  (b)  $W$  bosons emitted antiparallel
to the top quark spin are left-handed.}
\label{intuition}
\end{figure}

The distribution of the two helicity states of the $W$,
as viewed in the top rest frame is not uniform, but may be
intuitively understood from elementary
angular momentum conservation arguments and the $V{-}A$ coupling
between the $W$ and quarks.  
Since the mass of the $b$ quark is much smaller than the energy
imparted to it from the decay,
the left-handed chirality of the
$tbW$ vertex translates into a left-handed helicity for the $b$.
Suppose first that the $W$ boson is emitted along the top quark
spin axis, as in Fig.~\ref{intuition}a.  
Then, the spin of the $b$ points in the same direction
as the spin of the original $t$ and we must have 
zero spin projection for the $W$ boson ({\it i.e.}\ it must
be longitudinal).
On the other hand, when the $W$ is emitted in the
backwards direction (Fig.~\ref{intuition}b), the spin of the 
$b$ is opposite to the spin of the parent $t$.  
In this case, the $W$ must have left-handed
helicity in order to conserve angular momentum.

\begin{figure}[t]
\vskip4.7cm

\includegraphics{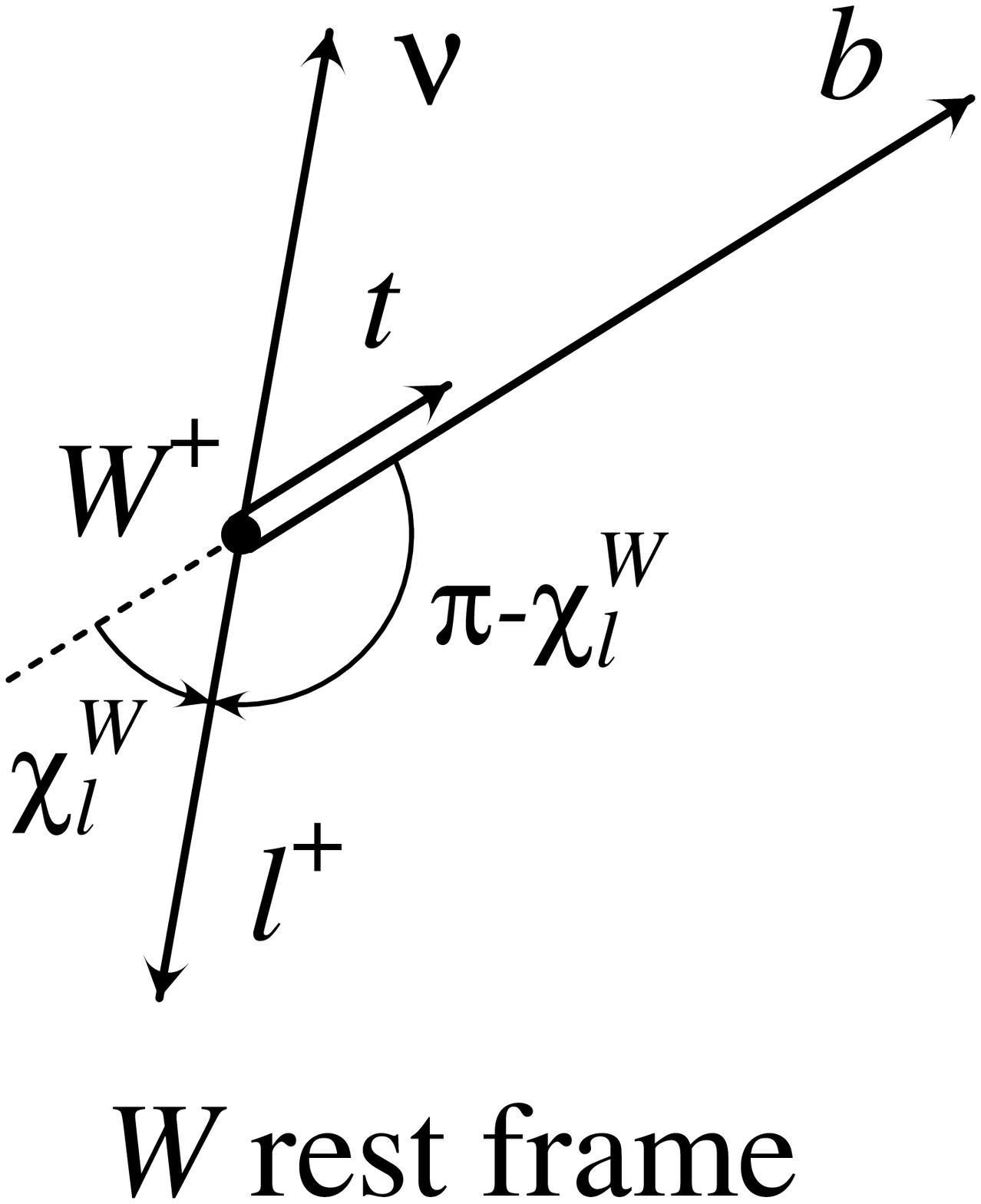}
\includegraphics{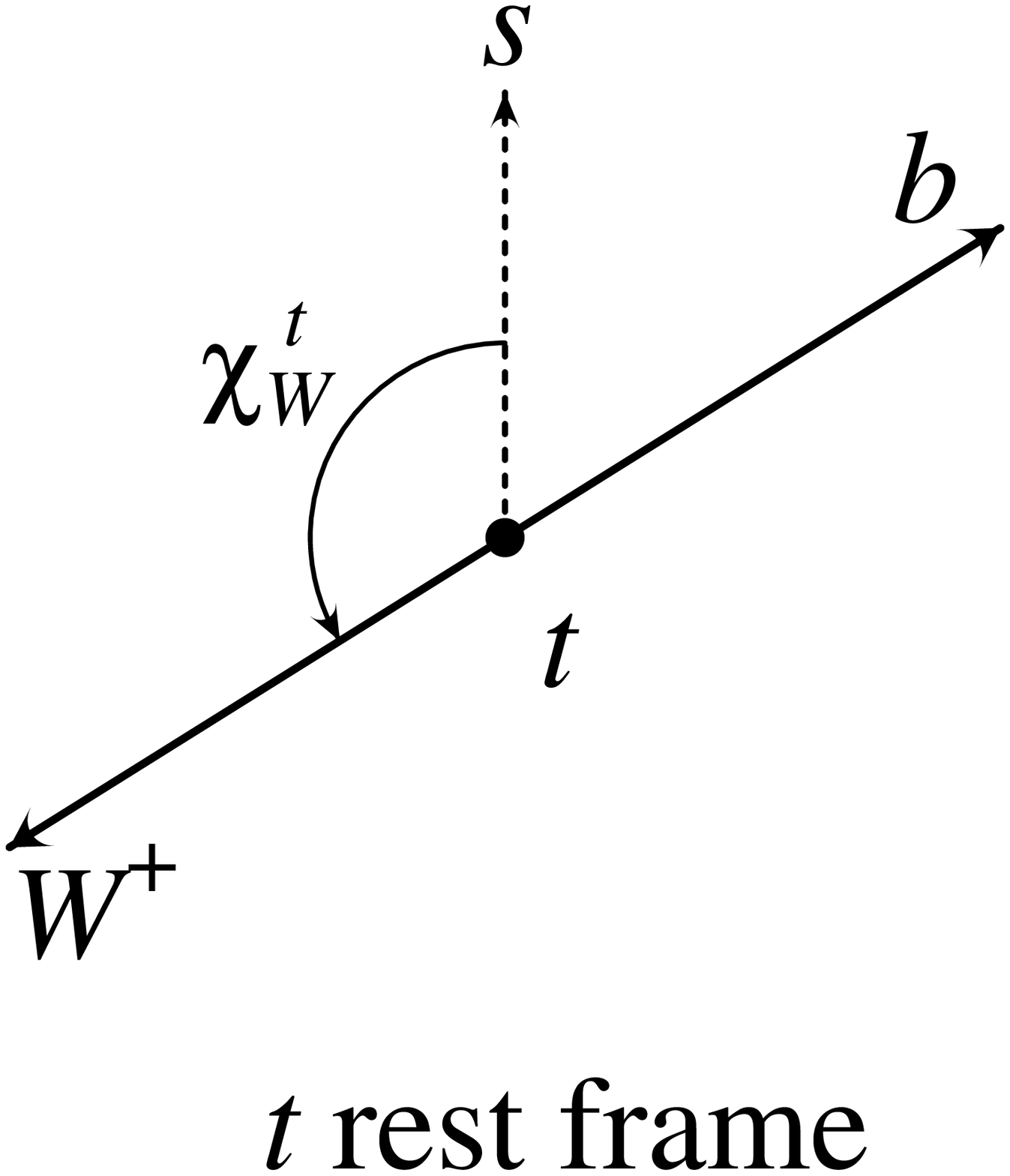}
\caption[]{Definition of the $W$ decay angles in top quark decay.
We define $\pi-\chi_\ell^{\W}$ to be the angle between the $b$
quark direction and the charged lepton direction in the $W$
rest frame.  The other interesting angle is $\chi_\W^{t}$,
the angle between the $W$ boson momentum and the top quark
spin axis in the top quark rest frame.
}
\label{Wdecay}
\end{figure}

\begin{figure}[t]
\vskip5.8cm
\includegraphics{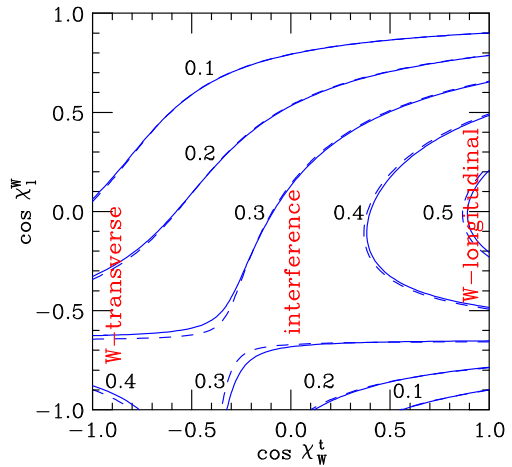}
\caption[]{Contours of the 
decay distribution of a spin up top quark 
in the $\cos\chi_\W^t$-$\cos\chi_\ell^\W$ plane.
$W$ bosons
emitted in the forward direction ($\cos\chi_\W^t\sim 1$)
are primarily longitudinal, while backward-emitted $W$'s
are mostly transverse.  Interference between the two spin
orientations dominates in the region around $\cos\chi_\W^t=0$.
The dashed lines indicate the distribution at 
tree level.\protect\cite{Winf}
The solid lines are the distribution including the ${\cal O}(\alpha_s)$
corrections.\protect\cite{tdecay-corrected}}
\label{W2D}
\end{figure}

Because the left-handed and longitudinal $W$'s are emitted
preferentially in different directions in the top quark
rest frame, there is an interesting correlation between
this angle ($\chi_\W^t$) and the emission angle ($\chi_\ell^\W$)
of the charged lepton
as viewed in the $W$ rest frame (see Fig.~\ref{Wdecay}).
This correlation is displayed in Fig.~\ref{W2D}, where the dashed
lines indicate the distribution at tree level\ts\cite{Winf}
and the solid lines show the consequences of including the
${\cal O}(\alpha_s)$ corrections to the decay matrix 
element.\cite{tdecay-corrected}  
Inclusion of the NLO QCD corrections does not greatly modify
the distribution:  the features we expected from our intuitive argument
are still clearly visible.
In particular, the $W$'s which
are emitted in the forward direction relative to the top spin 
show the characteristic $\sin^2\chi^\W_\ell$ associated with
their longitudinal polarization, whereas the backwards-going
$W$'s display the $\half(1-\cos\chi^\W_\ell)^2$ distribution
associated with their left-handed helicity.
%
%
\begin{figure}[t]
\vskip6.75cm
\includegraphics{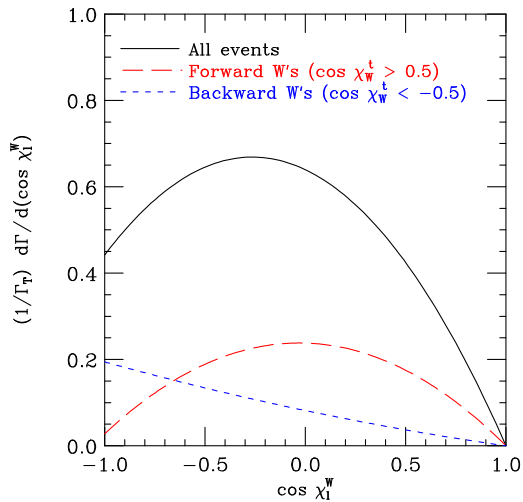}
\caption[]{Charged lepton angular distributions in the $W$
rest frame associated with the decay of a spin up top quark. 
Compared are samples selected on the basis of the emission angle
of the $W$ boson in the top quark rest frame.
}
\label{forback}
\end{figure}
Even if sufficient statistics to map out the complete
double differential distribution are not available, it may
still be possible to test this prediction.
In Fig.~\ref{forback} we divide the charged leptons into two
groups according to the direction of the parent $W$, and
plot the resulting distributions versus 
$\cos\chi_\ell^\W$.\footnote{In Fig.~\protect\ref{forback} 
we have arbitrarily chosen our cut point to be at 
$\cos\chi_\ell^\W = \pm 0.5$.  The optimal cut point for 
defining these 
two samples should be determined from a detailed
study including backgrounds and detector effects.}
The two samples clearly display differing distributions,
reflecting the correlation between
the emission angle of the $W$ boson and its helicity.
Of course, if 
we integrate the double differential distribution plotted
in Fig.~\ref{W2D} over all values of $\chi_\W^t$, we obtain the 
famous result for the distribution of the charged lepton in
the $W$ rest frame
\beq
{1 \over \Gamma_T}
{{d\Gamma}
\over
{d(\cos\chi_\ell^\W)}
}
= { 3\over4 } \thinspace
{ {\thinspace m_t^2 \sin^2 \chi_\ell^\W + 
 2m_\W^2 \hbox{$1\over2$}(1-\cos \chi_\ell^\W)^2}
\over
{m_t^2 + 2m_\W^2} },
\label{Wdist}
\eeq
which is
plotted as  the solid curve in Fig.~\ref{forback}.  Note that
the distribution in Eq.~(\ref{Wdist}) also applies to the 
decay of unpolarized top quarks, although the correlation
between $W$ polarization state and emission angle illustrated
in Figs.~\ref{Wdist} and~\ref{W2D} disappears in that case.

%


\section{Single Top Production}

Having described the features of the decays of polarized top
quarks, we now address the question of how to produce polarized
top quarks in the first place.  The answer is simple:  look
at events containing single top quarks.  Within the Standard
Model, there are three distinct channels for the production
of single tops, which we will now examine.

\begin{figure}[b]
\vskip3.8cm
\includegraphics{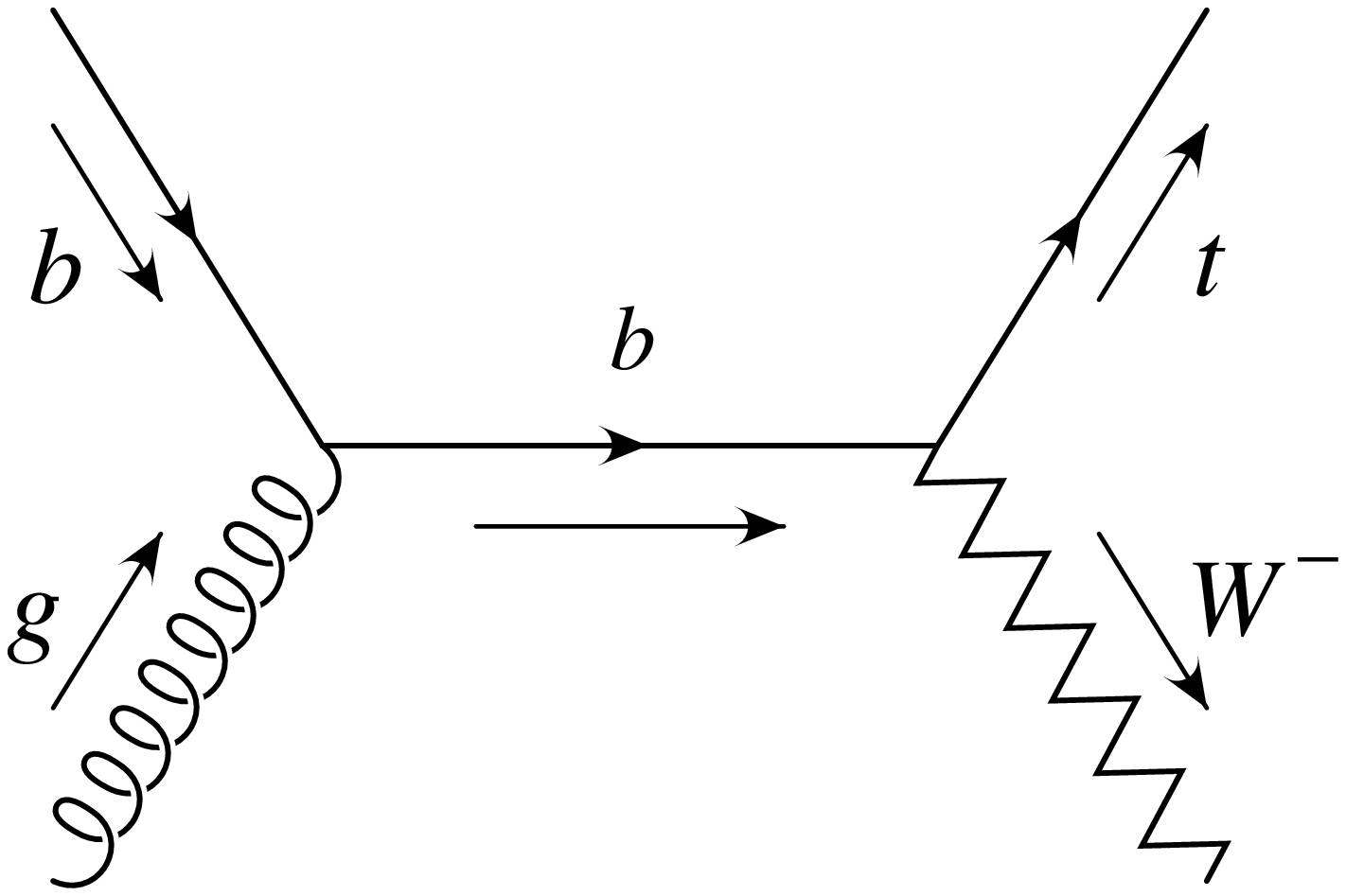}
\includegraphics{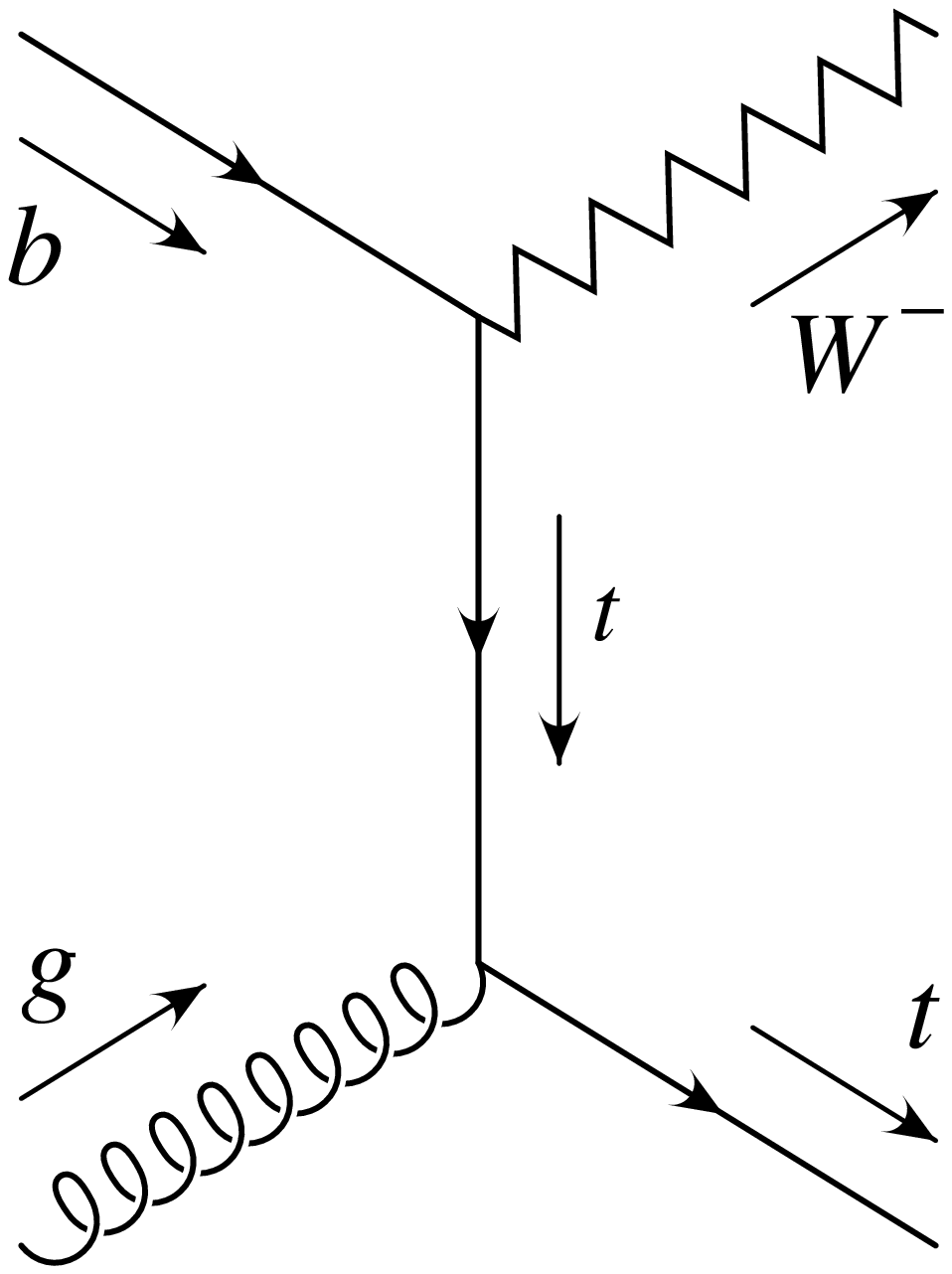}
\caption[]{Leading order Feynman diagrams for $tW^{-}$ associated
production.}
\label{tW}
\end{figure}

We begin our survey with associated 
production,$^{\ref{WtRefSTART}-\ref{WtRefEND}}$
where the top
quark is produced in conjunction with a $W$ boson.  The lowest order
graphs for this process are illustrated in Fig.~\ref{tW}.
Because this is a sea-sea process, the sum of $t$ and $\bar{t}$
production cross sections
is only of order 0.1 pb for $p\bar{p}$ collisions
at $\sqrt{s}=2\TeV$.  
This small production cross section when coupled to the
large insidious background from $t\bar{t}$ production
will make this process very difficult
to observe at Run II.\cite{WtRef3}  
The prospects for the observation of this channel are 
brighter at the LHC.\cite{LHCtop}  However, we have been
unable to construct a spin basis in which the tops 
in this channel are produced
with a high degree of polarization,
limiting the potential usefulness of this mode for
the study of angular correlations.


\begin{figure}[t]
\vskip2.7cm
\includegraphics{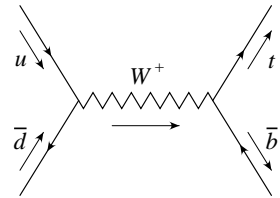}
\caption[]{Leading order
Feynman diagram for single top quark production in the
$W^{*}$ process.}
\label{WstarDiagram}
\end{figure}

The second-largest mode of single top production at the Tevatron
is predicted to be the $W^{*}$ or $s$-channel production
mechanism$^{\ref{WstarRefSTART}-\ref{WstarRefEND}}$
(see Fig.~\ref{WstarDiagram}).
Conceptually, this
is a very simple process:  a $u$ quark and a $\bar{d}$ quark
annihilate, forming  an off-shell $W$ which ``decays'' to a $t$ quark
and a $\bar{b}$ quark.  
Crossing symmetry trivially relates this diagram to the one 
for top decay followed by a hadronic $W$ decay.  Given that the
$d$-type quark is maximally correlated with the top quark spin,
it follows that in the production process, the top quark is 100\%
polarized in the direction of the $d$-type quark.\cite{OptimalBasis}
Since the proton beam
is a copious source of quarks and the antiproton beam a
copious source of antiquarks,
we intuitively expect that the $d$-type quark
will come from the antiproton beam most of the time.
In fact, we find that the antiproton beam supplies the $d$-type
quark more than 97\% of the time at a center of 
mass energy $\sqrt{s} = 2.0 \TeV$
(see Table~\ref{WstarProd}).
Thus, we define the {\it antiproton basis}\ to be that 
basis where the top quark spin is measured along the direction
of the antiproton beam momentum.\cite{OptimalBasis} 
For comparison, we also consider results
in the more traditional helicity basis.  Since helicity is
a frame-dependent concept for massive particles, the result
depends on whether we use the top quark direction in the
lab frame or zero momentum frame (ZMF) to define the helicity.
The spin contents in these three bases (lab helicity, ZMF helicity,
and antiproton) are listed in 
Table~\ref{WstarFractions}.
The third column of this table contains the
spin asymmetry
\beq
\Aud \equiv 
{ {N_\uparrow - N_\downarrow}
\over
  {N_\uparrow + N_\downarrow} }.
\label{Aud-def}
\eeq
This quantity governs the size of the observable angular
correlations:
in a situation where a mixture of
spin up and spin down top quarks is present,
Eq.~(\ref{dGamma}) becomes
\beq
{1\over\Gamma_T}\thinspace
{ {d\Gamma}\over{d(\cos\chi^t_i)} }
=
{1\over2}
\Bigl( 1+\Aud\alpha_i\cos\chi^t_i \Bigr).
\label{partial-polarization}
\eeq
From Table~\ref{WstarFractions} we see that the correlations
are predicted to be larger in the antiproton basis
by a factor of 1.7
over the lab frame helicity basis 
and by a factor of nearly 1.5
over the ZMF helicity basis.
Thus, the antiproton basis provides a significantly better
handle on the top quark spin than either of the two naturally-defined
helicity bases.

\begin{table}[t]
\centering
\caption{Fractional cross sections 
at leading order for single top quark
production in the $W^{*}$ channel at the Tevatron 
with $\protect\sqrt{s}= 2.0 \TeV$,
decomposed according to the parton content of the initial
state.\protect\cite{PDFs}
\lower2pt\hbox{\protect\phantom{j}}
\label{WstarProd}}

\begin{tabular}{cc@{\qquad}c@{\qquad}rc}
\multispan5\hrulefill \\[0.05cm]
\qquad\qquad\qquad\qquad
&   $p$     & $\bar{p}$ & Fraction & \qquad\qquad\qquad\qquad \\[0.1cm]
\multispan5\hrulefill \\[0.1cm]
&   $u$     & $\bar{d}$ & 97\phantom{.5}\% & \\
& $\bar{d}$ &   $u$     &  2\phantom{.5}\% & \\
&   $c$     & $\bar{s}$ & 0.5\% & \\
& $\bar{s}$ &   $c$     & 0.5\% & \\[0.1cm]
\multispan5\hrulefill 
\end{tabular}
%
%
\caption{Dominant spin fractions and asymmetries
at leading order
for single top quark production in the $W^{*}$
channel at the Tevatron with 
$\protect\sqrt{s} = 2.0 \TeV$.\protect\cite{PDFs}
\lower2pt\hbox{\protect\phantom{j}}
\label{WstarFractions}}

\begin{tabular}{ccccc}
\multispan5\hrulefill \\[0.05cm]
&Basis      & Spin Content & $\Aud$ &  \\[0.1cm]
\multispan5\hrulefill \\[0.1cm]
\ts\ts
\qquad\qquad &LAB helicity & 78\% L          & $-0.56$ & \qquad\qquad \\
             &ZMF helicity & 83\% L          & $-0.65$ & \\
             &antiproton   & 98\% $\Uparrow$ & $+0.96$ & \\[0.1cm]
\multispan5\hrulefill 
\end{tabular}
\end{table}

The ${\cal O}(\alpha_s)$ corrections to the total
rate for the $W^{*}$ process
have been computed by Smith and Willenbrock.\cite{Wstar1loop}
They are large, increasing the sum of the $t$ and $\bar{t}$
production cross sections from 0.59 pb to 0.88 pb ($+54\%$)
for the Tevatron at 2 TeV.  Clearly 
a calculation of this correction including the full 
top quark spin dependence is desirable to verify that the 
large top quark spin asymmetry in the antiproton basis 
continues to hold.


\begin{figure}[t]
\vskip6.7cm
\includegraphics{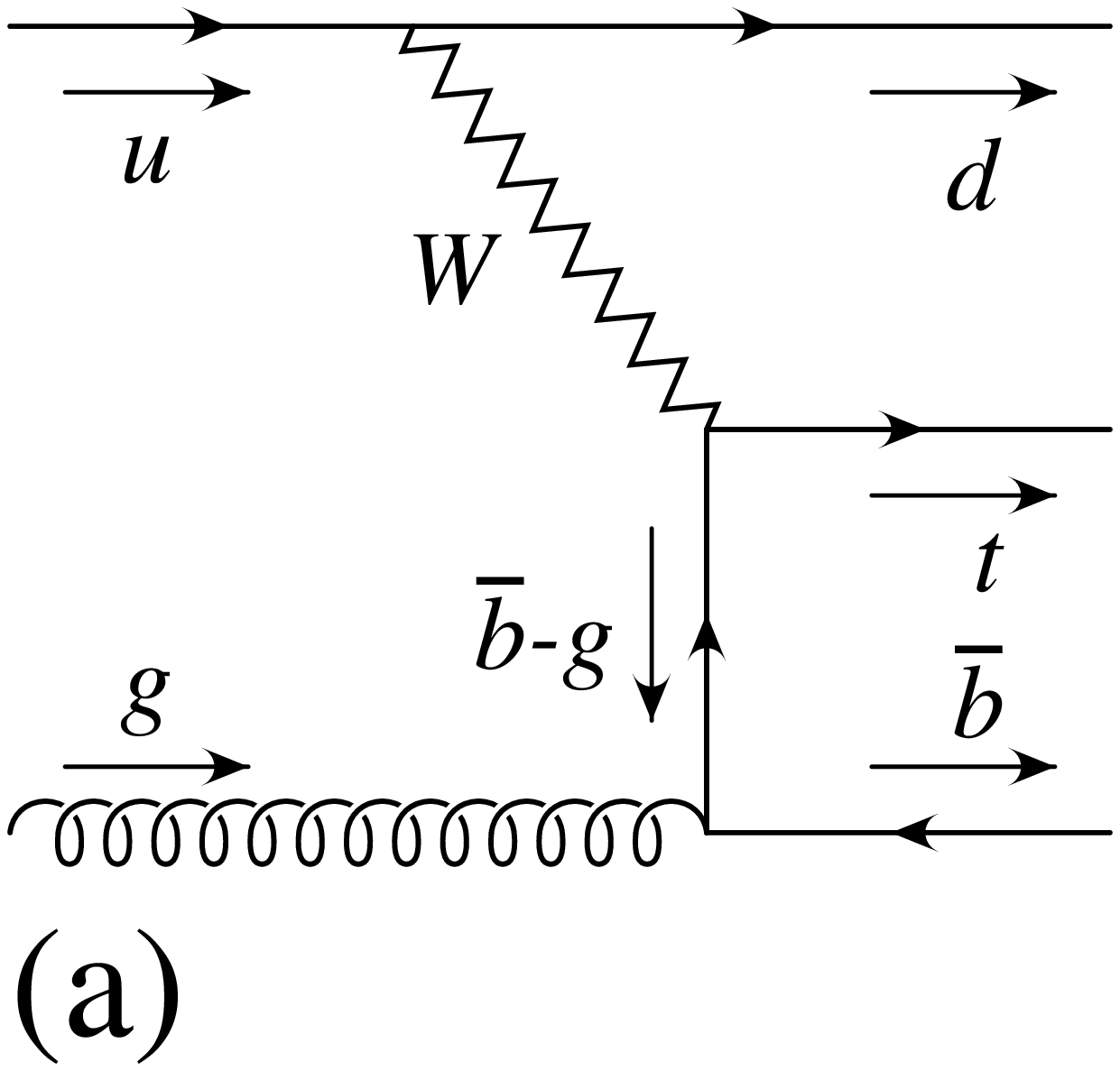}
\includegraphics{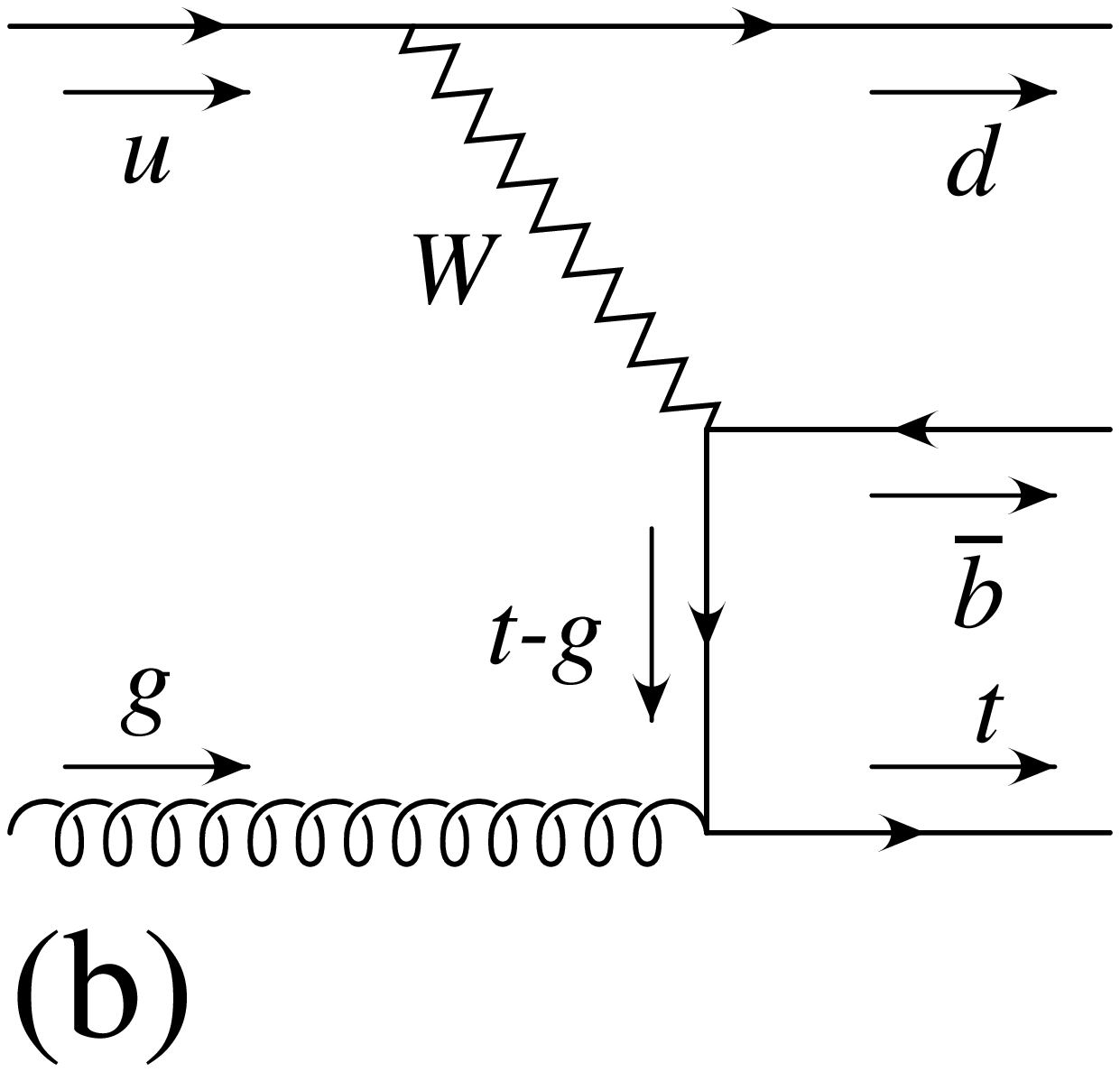}
\includegraphics{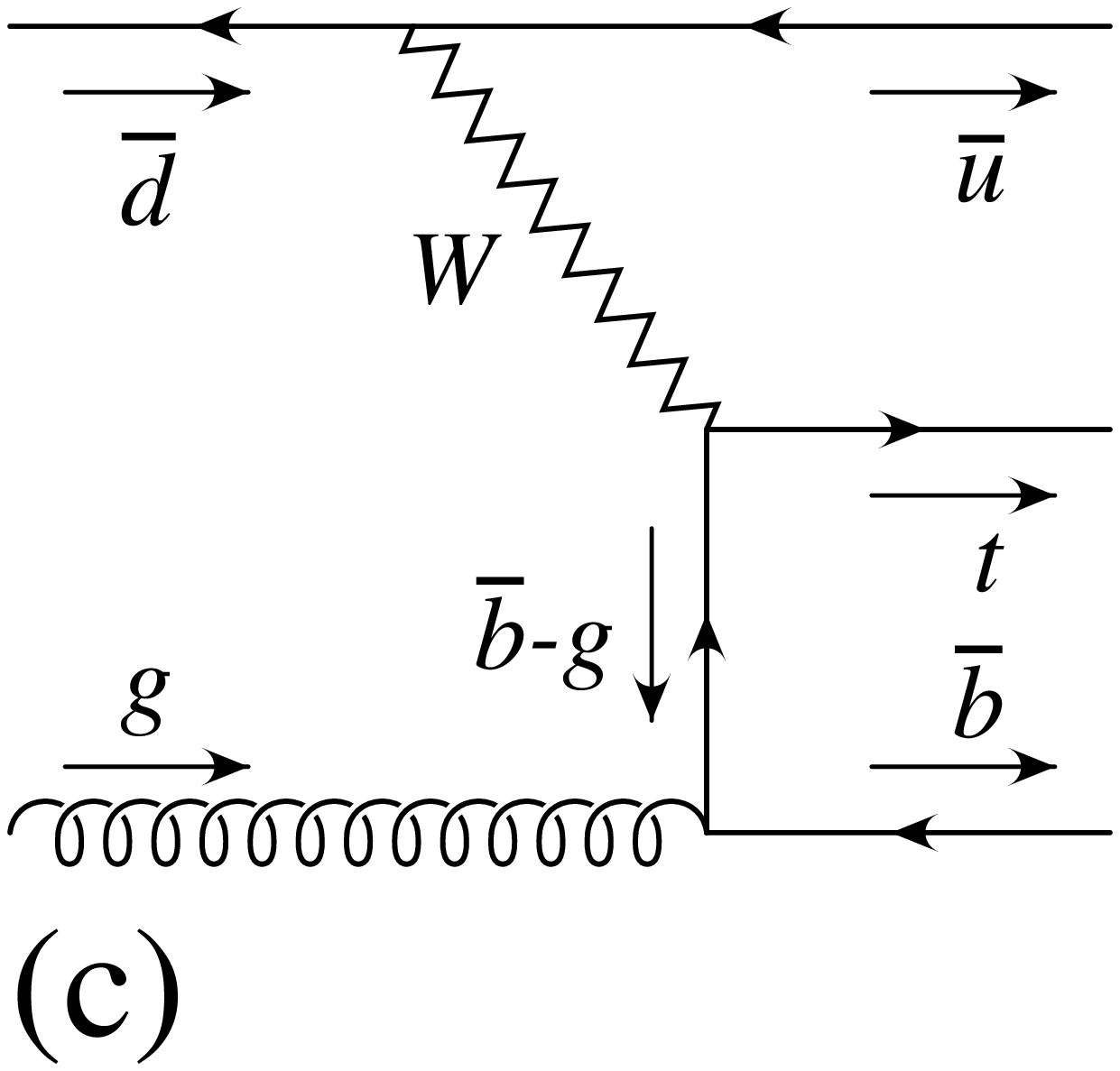}
\includegraphics{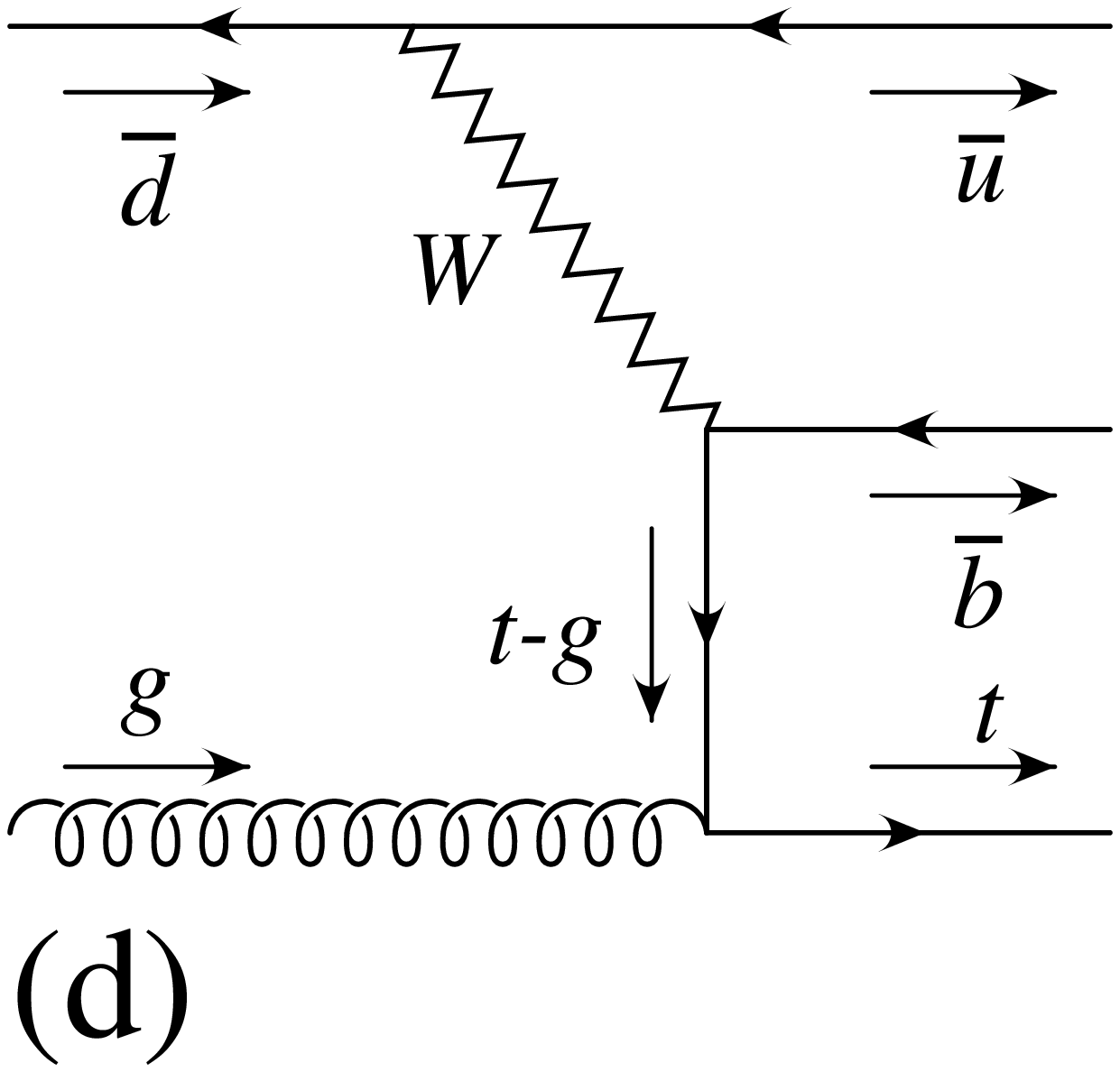}
\caption[]{Gauge-invariant set of Feynman diagrams for single
top quark production via $\Wg$-fusion.  The lower two diagrams
are related to the upper two diagrams by crossing symmetry.}
\label{WgDiagram}
\end{figure}

The third, and, in fact, dominant production mechanism for 
single top quarks
at the Tevatron at 2.0 TeV is the so-called $W$-gluon fusion
process,$^{\ref{WgRefSTART}-\ref{WgRefEND}}$
also referred to as the $t$-channel process.
At next-to-leading order, the sum of the $t$ and $\bar{t}$
cross sections in this mode is $2.12\pm0.10$ pb.\cite{ZacksTalk}
The lowest-order treatment of the spin correlations
in these events is based upon the two pairs of diagrams
illustrated in Fig.~\ref{WgDiagram}.\cite{OptimalBasis}
In these diagrams, a gluon from one beam
splits into a $b\bar{b}$ or $t\bar{t}$ pair,
which fuses with a $W$ radiated from a light ($u$ or $\bar{d}$\ts)
spectator quark from the other beam. 
The final state contains a (relatively low-$p_T$)
$\bar{b}$ jet, the top quark
decay products, and the spectator jet.

The production of top quarks whose spin is {\it anti-aligned}\ with
the direction of the $d$-type quark proceeds entirely through
diagrams (b) and (d), where the gluon splits into a $t\bar{t}$ pair.
On the other hand,
top quarks whose spin is aligned with the $d$-type quark directions
come from all four diagrams.  Diagrams (a) and (c) are divergent
as $m_b \rightarrow 0$, and dominate the lowest-order result
for $m_b \ll m_t$.  Hence, the top quark spin is highly correlated
with the direction of the $d$-type quark, being aligned with 
this axis $97\%$ of the time.\cite{OptimalBasis} 

Of course, it is not possible in any given event to know where
the $d$-type quark is with certainty.  In fact, the $d$-type
quark could be contained in either beam or in the spectator jet.
We know that the $u$ quark content  of the proton is greater
than the $\bar{d}$ quark content of the antiproton.
Furthermore, the gluon content of both is the same.
Hence, we expect that the largest share of the cross
section comes from $ug \rightarrow t\bar{b}d$, with
the spectator jet containing the $d$ quark.  As we can
see from Table~\ref{WgProd}, this expectation is correct.
In fact, the spectator jet contains the $d$-type quark approximately
three-quarters
of the time.
Furthermore, in those events where the $d$-type quark is
in the initial state, the fact that the spectator jet tends
to be produced in the forward direction means that it is
still not a bad choice for the spin quantization axis:  it is
``almost'' in the ideal direction.
Thus, we define the {\it spectator basis}\ as the basis
in which we choose the spin axis to be aligned with the
momentum of the spectator jet.\cite{OptimalBasis}
In this basis, the top
quark is produced in the spin up state 95\% of the time
at leading order
(see Table~\ref{WgFractions}).

\begin{table}
\centering
\caption{Fractional cross sections for single top quark
production in the $\Wg$-fusion channel at the Tevatron 
with $\protect\sqrt{s} = 2.0 \TeV$,
decomposed according to the parton content of the initial
state\ts\protect\cite{PDFs} ($2\rightarrow3$ diagrams only).
\lower2pt\hbox{\protect\phantom{j}}
\label{WgProd}}

\begin{tabular}{cc@{\qquad}c@{\qquad}rc}
\multispan5\hrulefill \\[0.05cm]
\qquad\qquad\qquad\qquad
&   $p$     & $\bar{p}$ & Fraction & \qquad\qquad\qquad\qquad \\[0.1cm]
\multispan5\hrulefill \\[0.1cm]
&   $u$     &   $g$     & 67\% & \\
&   $g$     &   $u$     &  3\% & \\
&   $c$     &   $g$     &  1\% & \\
&   $g$     &   $c$     &  1\% & \\[0.1cm]
\multispan5\hrulefill \\[0.1cm]
&   $g$     & $\bar{d}$ & 21\% & \\
& $\bar{d}$ &   $g$     &  3\% & \\
&   $g$     & $\bar{s}$ &  2\% & \\
& $\bar{s}$ &   $g$     &  2\% & \\[0.1cm]
\multispan5\hrulefill 
\end{tabular}
\end{table}
\begin{table}
\centering
\caption{Dominant spin fractions and asymmetries 
for single top quark production in the $\Wg$-fusion
channel at the Tevatron with 
$\protect\sqrt{s} =  2.0 \TeV$.\protect\cite{PDFs}
Compared are the leading order result ($2\rightarrow3$ diagrams
only) and the leading order plus resummed large logs result
(the contributions represented in Fig.~\protect\ref{resummed}).
\lower2pt\hbox{\protect\phantom{j}}
\label{WgFractions}}

\begin{tabular}{cccrccr}
\multispan7\hrulefill \\[0.05cm]
   &&\multispan2 $2{\rightarrow}3$ only 
   && \multispan2 \nts LO$+$large\ts\ts logs\ns \\[0.1cm]
Basis && Spins & $\Aud$ && Spins & $\Aud$\\[0.1cm]
\multispan7\hrulefill \\[0.1cm]
LAB helicity  &&   64\% L  & $-0.28$ && 68\% L & $-0.36$ \\
ZMF helicity  &&   83\% L  & $-0.65$ &&  \multispan2 undefined \\
spectator     &&   95\% $\Uparrow$ & $+0.90$ && 
                  96\% $\Uparrow$ & $+0.93$            \\[0.1cm]
\multispan7\hrulefill 
\end{tabular}
\end{table}

\begin{figure}[t]
\vskip12.0cm
\includegraphics{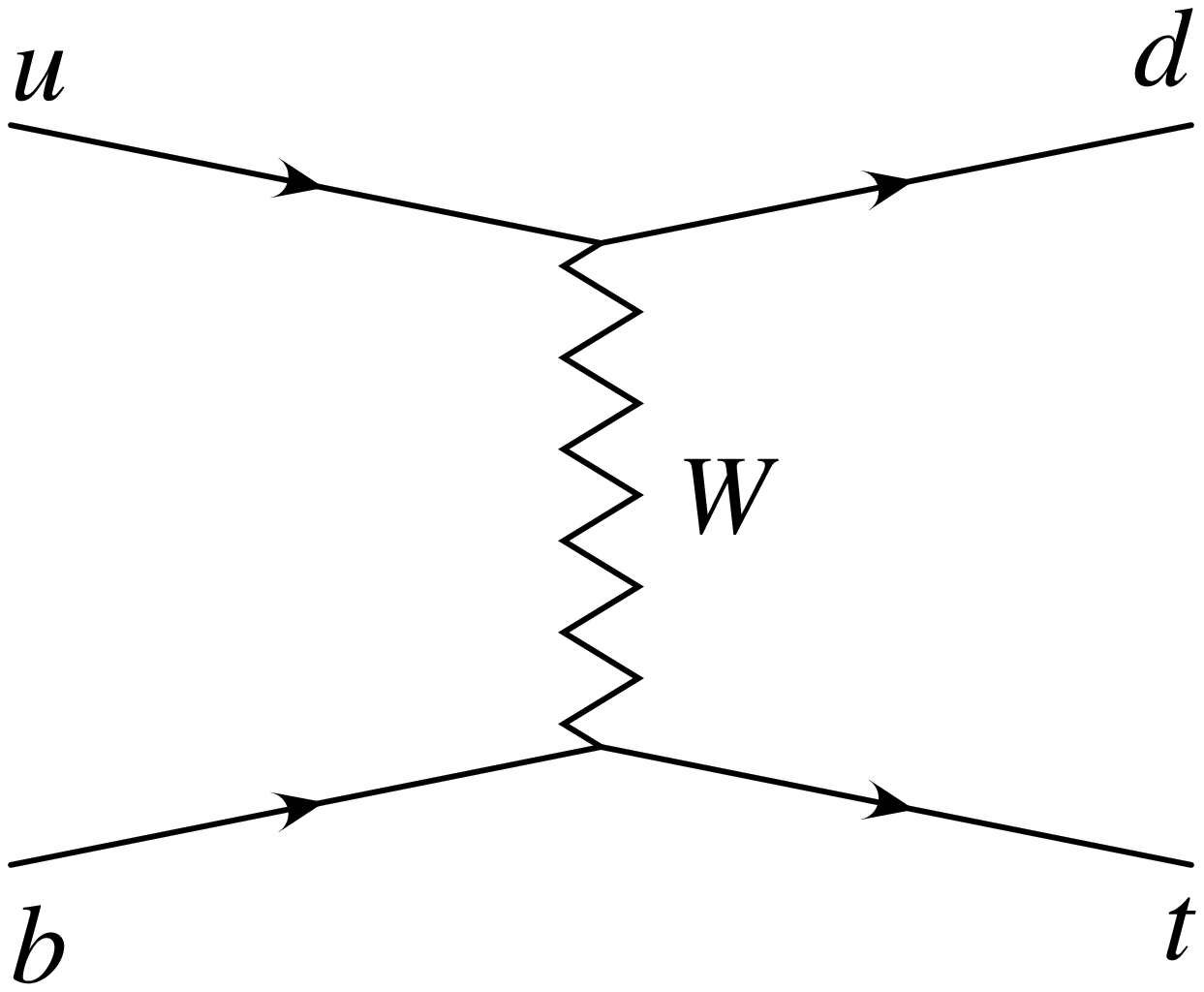}
\includegraphics{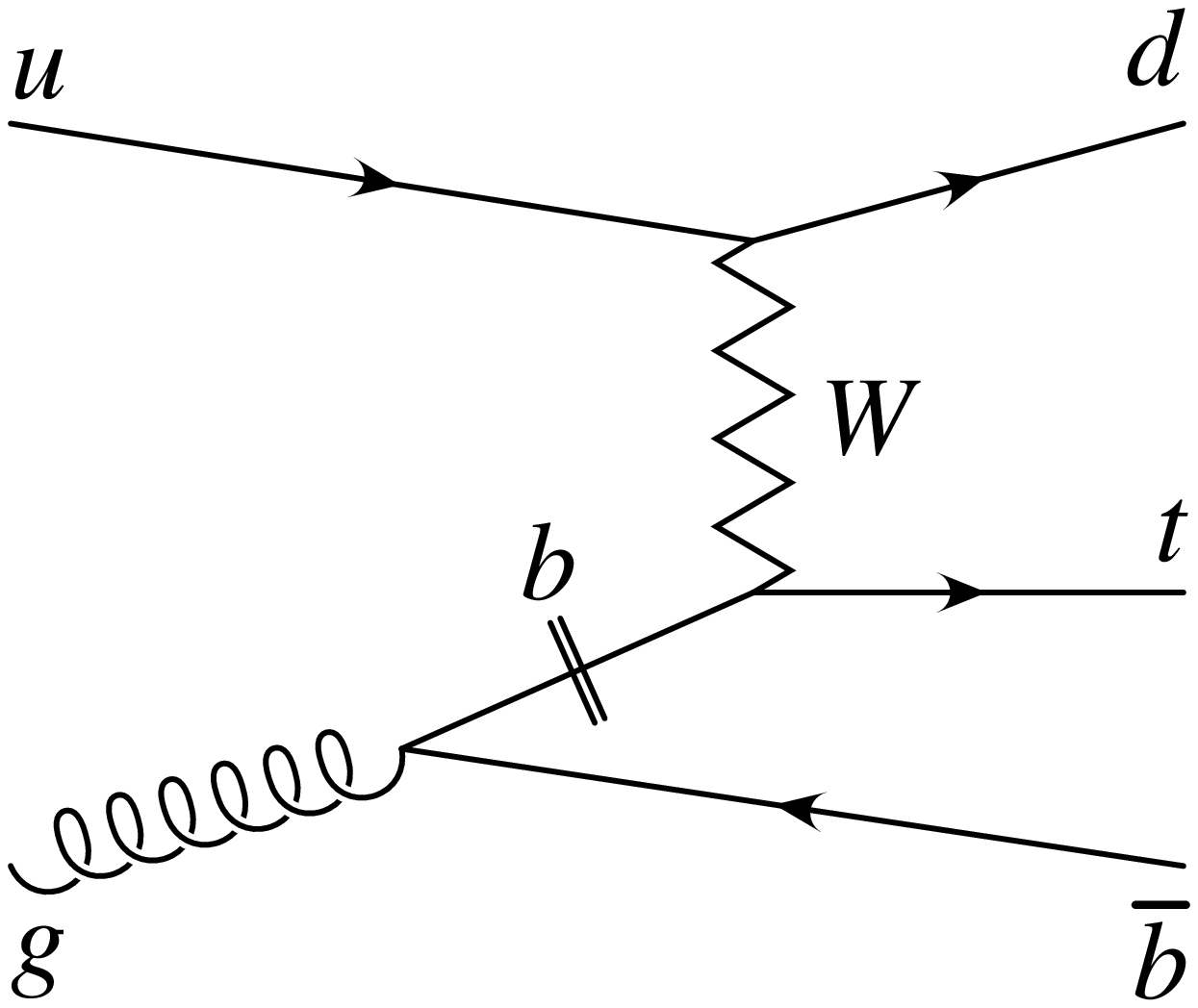}
\includegraphics{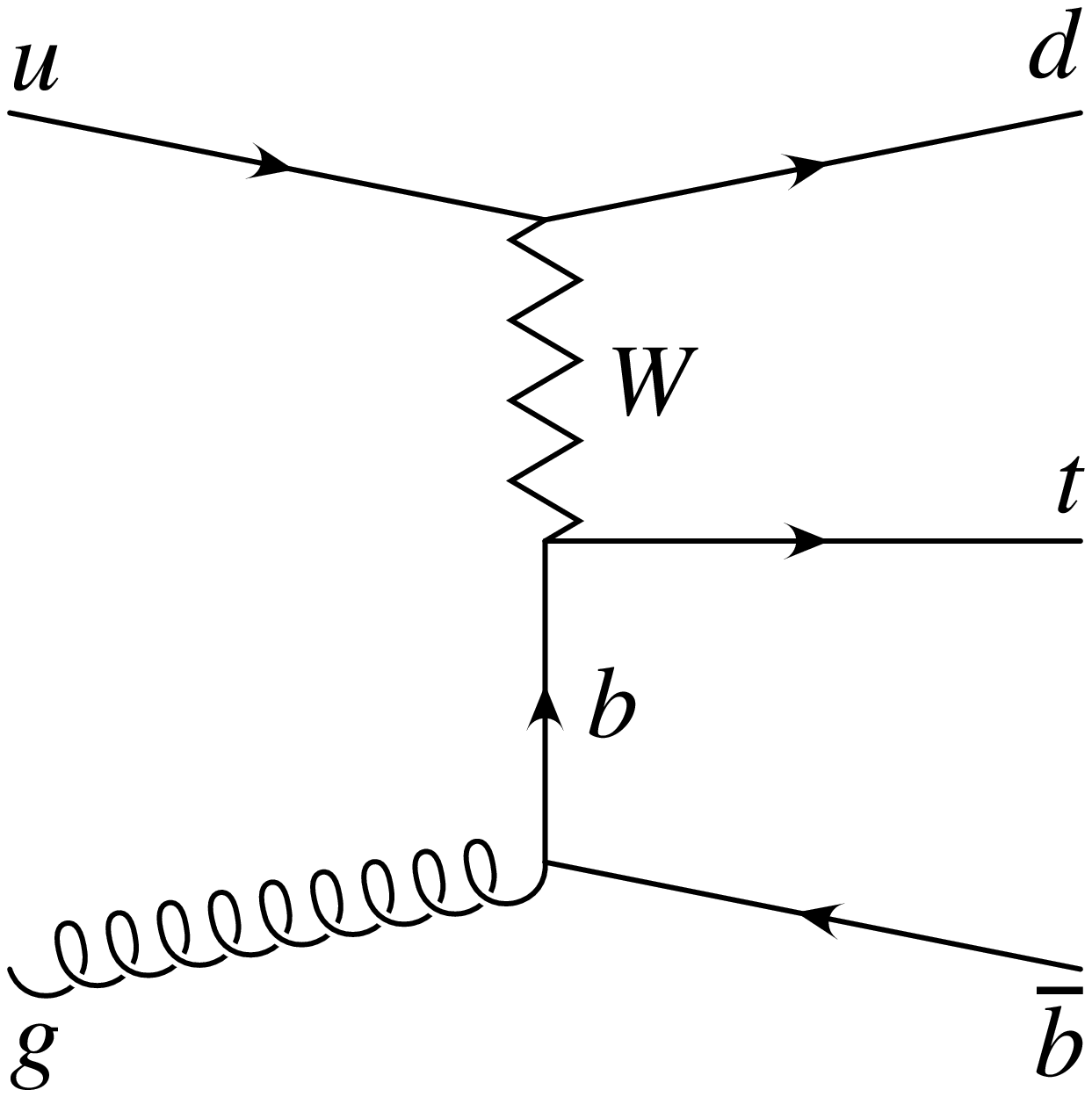}
\includegraphics{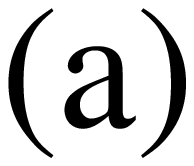}
\includegraphics{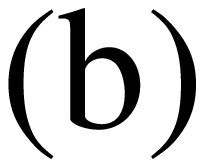}
\includegraphics{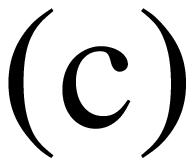}
\includegraphics{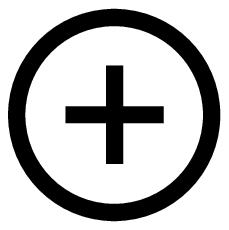}
\includegraphics{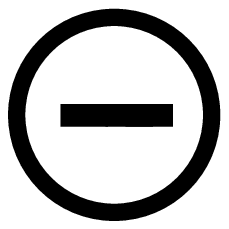}
\includegraphics{Figs/plus.ps}
\caption[]{Representative Feynman diagrams for 
for the leading order plus resummed large logs computation
of single top quark production via $\Wg$-fusion.  
The middle diagram represents the overlap between the $2\rightarrow2$
and $2\rightarrow3$ processes which results when the
powers of $\ln(m_t^2/m_b^2)$ are resummed and absorbed into
the $b$-quark parton distribution function.}
\label{resummed}
\end{figure}

The situation at next-to-leading order is somewhat complicated.
The divergence of diagrams (a) and (c) as $m_b\rightarrow 0$
leads to large logarithms of the form $\ln(m_t^2/m_b^2)$.
Because these logarithms convert the perturbation series from
an expansion in $\alpha_s$ to one in $\alpha_s\ln(m_t^2/m_b^2)$,
it is desirable to resum these logs by introducing a parton
distribution function for the $b$ 
quark.$^{\ref{WgRefEND},\ref{HQpdfSTART}-\ref{HQpdfEND}}$
Once we have done this,
the leading-order diagrams are for the $2\rightarrow 2$ process
represented in Fig.~\ref{resummed}a.  The original $2\rightarrow 3$
process (Fig.~\ref{resummed}c) becomes subleading. 
Because the logarithmic portion of the $2\rightarrow3$ process
is already included in the $b$ quark PDF used to compute the 
$2\rightarrow 2$ process, it is necessary to subtract the
overlapping portion of these two diagrams to avoid double
counting (Fig.~\ref{resummed}b).
The combination of the three diagrams in Fig.~\ref{resummed}
may be described as ``leading order plus resummed large logs.''
To complete the full NLO computation, we must add in the 
``true'' $\alpha_s$ corrections coming from loops and
soft parton radiation.  

The spin structure of the three diagrams in Fig.~\ref{resummed}
is understood.\cite{WgLHC} 
The $2\rightarrow3$ diagrams give precisely the
same contributions as were described above:
the top quark spin is aligned with the $d$-type quark 97\%
of the time.  The $2\rightarrow2$ diagram, 
being a crossed version of $t$ decay, produces tops whose
spin is aligned with the $d$-type quark 100\% of the time.
Because the overlap piece is to be evaluated with the intermediate
$b$ quark on-shell, the overlap term also has the same spin 
structure as the $2\rightarrow2$ piece.
As a consequence of considering the next-to-leading order process,
it becomes impossible to unambiguously define what is meant by
the ZMF helicity basis, even theoretically.  
That is, should we take the ZMF of the initial
state or the ZMF of the final state?  
Experimental considerations do not resolve this ambiguity,
since on one hand
we cannot directly measure the momentum fractions 
of the incoming partons, and on the other hand the $\bar{b}$
quark is hard to see.  Thus, in reality, there are only two viable
choices for the spin basis:  the lab frame helicity basis and the
spectator basis.  As we can see from the final two columns of 
Table~\ref{WgFractions}, the spectator basis is superior to
the lab frame helicity basis, producing correlations which are
larger by a factor of 2.6.
Not included in Table~\ref{WgFractions} is the effect of the
``true'' ${\cal O}(\alpha_s)$ corrections,
since their spin dependence has not yet been computed.
These corrections
shift the total cross section by approximately 12\%.\cite{Stelzer1}

\section{Observing Correlations}

We now briefly address the question of actually trying to
observe these correlations in a collider environment
such as the Tevatron.\footnote{Studies of the situation at the LHC
appear in Sec.~5.4 of Ref.~\protect\ref{LHCtop}.}
Consider single top quarks generated by the $\Wg$-fusion process.
We know that if the direction of the spectator jet is chosen
as the spin quantization axis, then $\Aud = 0.93$.
Furthermore, if we choose those events where the
$W$ boson decays leptonically, then the charged lepton
has the strongest correlation with the top quark spin,
$\alpha_\ell = 1$.  Hence, the most distinctive
correlations appear when we measure the angle $\chi_\ell^t$
between the spectator jet
and the charged lepton in the rest frame of the top quark.

\begin{figure}[t]
\vskip5.9cm
\includegraphics{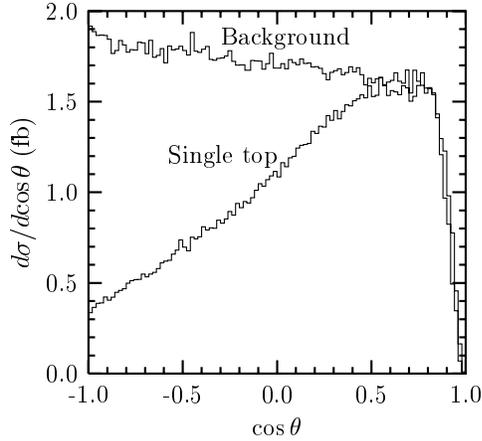}
\caption[]{Monte Carlo results of 
Stelzer, {\it et al.}\ts\protect\cite{Study}, for the
top quark rest frame
angular distribution of the charged lepton in single
top quark events at the Tevatron ($\protect\sqrt{s}=2 \TeV$),
with respect to the untagged jet. 
Also shown is the angular distribution of the background
events passing their selection criteria.} 
\label{StelzerFigure}
\end{figure}

Stelzer {\it et al.}
have performed a first study of how well this measurement
can be performed at the Tevatron.\cite{Study}
They simulate the $\Wg$-fusion
process at 2~TeV including backgrounds and the effects of
energy smearing, jet reconstruction, and cuts
(for complete details, see Ref.~\ref{Study}).
Their data set for the study of spin correlations
consists of events where there is exactly one $b$-tagged and
one untagged jet with a transverse momentum above 20~GeV.
These cuts help to control the background from $t\bar{t}$.
The untagged jet is assumed to be the spectator jet.
The top quark rest frame is reconstructed by selecting 
the solution for
the neutrino momentum with the smallest magnitude of rapidity.
Using these assumptions, they plot the distribution
in the angle $\theta$, which is the angle between the
charged lepton and the untagged jet in the reconstructed
top quark rest frame.  This angle is their best approximation to
the theoretical angle $\chi_\ell^t$.
Their results are shown in Fig.~\ref{StelzerFigure}.  
The bins near $\cos\theta = 1$ are depleted by the
isolation cut imposed on the lepton to distinguish it from
the spectator jet.  Away from this region, the signal
shows a distinct slope, while the background is nearly flat.

To quantify the size of the correlations
present in Fig.~\ref{StelzerFigure},
Stelzer {\it et al.}\ 
define a
cross section asymmetry
over the range $-1 \leq \cos\theta \leq 0.8$, excluding
the region where the cuts are most troublesome.
Specifically,
\beq
A \equiv
{
{ \sigma(-1{\le}\cos\theta{\le}-0.1) 
  - \sigma(-0.1{\le}\cos\theta{\le}0.8) }
\over
{ \sigma(-1{\le}\cos\theta{\le}-0.1) 
  + \sigma(-0.1{\le}\cos\theta{\le}0.8) }
}.
\eeq
In the ideal case where we
assume that we correctly identify the spectator jet 100\%
of the time and measure all momenta and angles (including
those associated with the neutrino) perfectly,
then we would obtain $A=-45\%$.
The authors of Ref.~\ref{Study} report that for the signal
distribution in Fig.~\ref{StelzerFigure}, $A=-38\%$.  The difference
is attributable to the effects of cuts, smearing, 
and the use of $\theta$ instead of $\chi_\ell^t$.
Of course, what would
be measured in a real experiment is the sum of the signal and
background distributions, in which case $A$ is further reduced
to $-14\%$.  Nevertheless, with the 2 fb$^{-1}$ of integrated
luminosity expected at Run IIa, this asymmetry should be visible at
the $3\sigma$ level.\cite{Study}
The $5\sigma$ level requires a total
of 5 fb$^{-1}$, which should be easily reached by Run  IIb.

\section{Summary}

The Standard Model weak decays of polarized top quarks have
a rich structure of angular correlations induced by the
$V-A$ structure of the $Wtb$ vertex.  The charged lepton or
$d$-type quark from the decaying $W$ boson is maximally correlated
with the spin of the parent top quark.  
Interference between the left-handed and longitudinal
$W$ bosons plays a crucial role in producing this correlation.
The emission angle of the $W$ boson with respect to the top
quark spin is correlated with its polarization state.
Longitudinal (left-handed) $W$ bosons are emitted 
primarily parallel (antiparallel) to the direction of
the top quark spin in its rest frame.

Single top production provides a source of polarized top
quarks.  With the exception of the associated production mode
(which is too small to observe at the Tevatron), single top
quarks are produced with a large degree of polarization
when the appropriate spin basis is chosen.
In the antiproton basis, 98\% of the top quarks produced
in the $W^{*}$ channel have spin up.  For $\Wg$-fusion,
96\% of the top quarks are produced with spin up in the
spectator basis.  The NLO corrections to the $W^{*}$ mode
are not known in a spin-dependent form.  
The portion of the
higher order corrections which resums the large logarithms
$\ln(m_t^2/m_b^2)$ are known including
the spin dependence in the $\Wg$-fusion case.  They serve
to increase the net polarization of the top quark 
by a small amount over the leading order result.

A recent study
has shown for the $\Wg$-fusion mode that 
the correlation between the top quark spin axis and the
charged lepton
should be observable at the $3\sigma$ level in Tevatron Run IIa.
Additional studies are required to determine which of the
other correlations associated with polarized top decay will
be observable at the Tevatron.

\section*{Acknowledgments}

Special thanks to the Thinkshop$^2$ organizers and especially
the organizers of the Weak Interactions discussion group
for a stimulating program.
Thanks to Tim Stelzer, Zack Sullivan, and
Scott Willenbrock for the use of their figure from Ref.~\ref{Study}
in this talk.

\medskip\noindent
High energy physics research at McGill University is
supported in part by the Natural Sciences and Engineering 
Research Council of Canada and the Fonds pour la Formation de 
Chercheurs et l'Aide \`a la Recherche of Qu\'ebec.
I would also like to thank the Fermilab theory group for
their hospitality and support during my visit.


\end{document}